\documentclass[
 reprint,
superscriptaddress,
nofootinbib,
nobibnotes,
 amsmath,amssymb,
prl,footinbib
]{revtex4-1}
\usepackage{graphicx}
\usepackage{dcolumn}
\usepackage{bm}
\usepackage{notoccite}
\usepackage{physics}
\usepackage{xcolor}
\usepackage{mathtools}

\begin{document}

\title{Berry Curvature Dipole in Strained Graphene: a Fermi Surface Warping Effect}

\author{Raffaele Battilomo}
\affiliation{Institute for Theoretical Physics, Center for Extreme Matter and Emergent Phenomena, Utrecht University, Princetonplein 5, 3584 CC Utrecht,  Netherlands}
\author{Niccol\'o Scopigno}
\affiliation{Institute for Theoretical Physics, Center for Extreme Matter and Emergent Phenomena, Utrecht University, Princetonplein 5, 3584 CC Utrecht,  Netherlands}
\author{Carmine Ortix}
\affiliation{Institute for Theoretical Physics, Center for Extreme Matter and Emergent Phenomena, Utrecht University, Princetonplein 5, 3584 CC Utrecht,  Netherlands}
\affiliation{Dipartimento di Fisica ``E. R. Caianiello", Universit\'a di Salerno, IT-84084 Fisciano, Italy}

\begin{abstract}
It has been recently established that optoelectronic and non-linear transport experiments can give direct access to the dipole moment of the Berry curvature in non-magnetic and non-centrosymmetric materials. 
Thus far, non-vanishing Berry curvature dipoles have been shown to exist in materials with substantial spin-orbit coupling where low-energy Dirac quasiparticles form tilted cones. 
Here, we prove that this topological effect does emerge in two-dimensional Dirac materials even in the complete absence of spin-orbit coupling. In these systems, it is the warping of the Fermi surface that triggers sizeable Berry dipoles. We show indeed that uniaxially strained monolayer and bilayer graphene, with substrate-induced and gate-induced band gaps respectively, are characterized by Berry curvature dipoles comparable in strength to those observed in monolayer and bilayer transition metal dichalcogenides.
\end{abstract}

\maketitle

\paragraph{Introduction --}
Because of their deep relation to topology~\citep{Qi2011,KaneHasanrev} the family of Hall effects~\citep{Niurev,AHE}, the most famous member of which is the quantization of the Hall conductance in strong magnetic fields~\citep{VonKlitzing1980,Thouless,Niu}, have been intensively scrutinized in recent years. A prerequisite for any Hall effect to appear is time-reversal symmetry breaking. Therefore, either magnetic fields or magnetic dopants are required to have a non-vanishing Hall conductance. 
It has been recently established, however, that a Hall-like current can still be observed in non-centrosymmetric systems as a non-linear response to an external electric field \citep{Sodemann2015,Moore}. Such non-linear Hall current is closely related to the Berry curvature dipole, which is essentially the first moment of the Berry curvature in momentum space. As a result, this non-linear effect can be used as a direct probe of the  geometry of the Bloch states in time-reversal invariant systems.

 Single-layer transition metal dichalcogenides MX$_2$ (M=Mo,W and X=S, Se, Te) \citep{TMD} have been theoretically predicted \citep{TMD2,Binghai} and experimentally verified \citep{Son,Xu2018,Kang} as material platforms supporting large Berry curvature dipoles. These materials possess large spin-orbit coupling and lack of inversion symmetry in their $T_d$ structure. Furthermore, inversion symmetry breaking can be also achieved with the aid of external electric fields in the topological non-trivial $1 T^{\prime}$ phase. A non-linear Hall effect has been also predicted \citep{Du} and experimentally observed in bilayer WTe$_2$ \citep{Ma2019}. Finally, different non-centrosymmetric three-dimensional materials have been proposed to feature sizable Berry curvature dipoles. These include the topological crystalline insulator SnTe \citep{TCI}, which undergoes a ferroelectric distorsion at low temperatures \citep{Lau}, time-reversal symmetric Weyl semimetals in the TaAs materials class \citep{Sodemann2015}, as well as the giant Rashba material BiTeI \citep{Facio2018}.  

\begin{figure}[tbp]
\includegraphics[width=1\columnwidth]{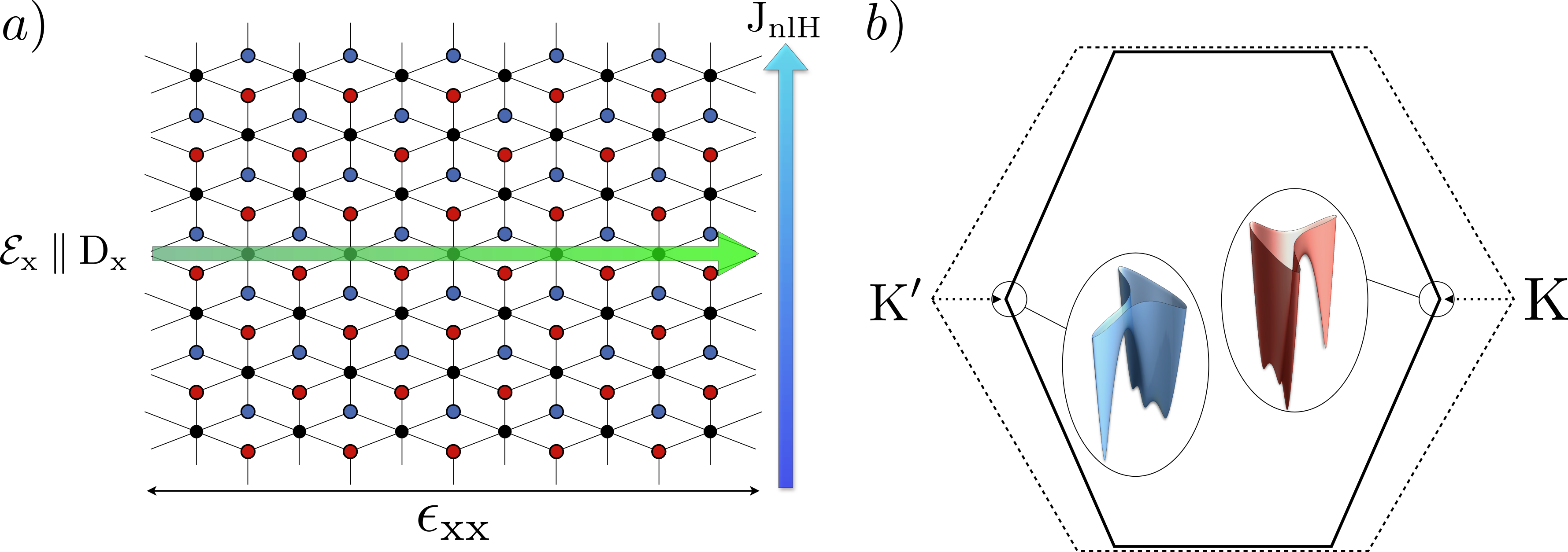}
\caption{a) Top view of the uniformly strained Bernal-stacked bilayer graphene lattice. Red and blue carbon atoms correspond to different layer and are not equivalent due to the applied voltage that breaks inversion symmetry. Black sites indicate the overlap between atoms of the two different layers. By applying an electric field  $\mathcal{E}_x$ parallel to the dipole $\mathrm{D_x}$, a non-linear Hall current $\mathrm{J_{nlH}}$ is generated. b) Strain deformation of the Brillouin zone. The warped low-energy dispersion is shown in the vicinity of the two unstrained high symmetry points $K$ and $K'$.}
\label{fig:schematic}
\end{figure}

The common trait of all these materials is their strong spin-orbit coupling and the presence of low-energy Dirac quasi-particles forming tilted Dirac cones. In particular, the tilt of the Dirac cones does not change the Berry curvature of the system but it is crucial to get a corresponding non-vanishing first dipole moment. 
The aim of this work is to show that this topological effect emerges even in the complete absence of spin-orbit coupling in two-dimensional Dirac materials. In these systems, the non-vanishing Berry curvature dipole does not stem from the presence of tilted Dirac cones but it is due to the warping of the Fermi surface.
We show indeed that a sizable Berry curvature dipole arises in uniaxially strained single-layer and bilayer-graphene where inversion symmetry is broken by the existence of a substrate and an external electric field, respectively. In the absence of shear strains, the non-vanishing dipole in these materials is generated along the zig-zag direction, which is orthogonal to the armchair mirror line (Fig. \ref{fig:schematic}a). Even more importantly, the appearance of a finite dipole can only be captured taking explicitly into account the terms accounting for the warping of the Fermi surface (Fig. \ref{fig:schematic}b) in the low-energy description of the system. The warping-induced Berry dipole is strongly enhanced in Bernal-stacked bilayer graphene and reaches the nanometer scale, which is comparable to the value experimentally observed in bilayer WTe$_2$~\cite{Ma2019}.

\paragraph{Berry Curvature Dipole in Strained Monolayer Graphene --} 
We start out by recalling the relation between the non-linear Hall current and the Berry curvature dipole as it can be derived using the semiclassical Boltzmann picture of transport \citep{Sodemann2015}. In time-reversal invariant, non-centrosymmetric crystals applying an AC electric field $E_c=Re(\mathcal{E}_c e^{i\omega t})$ induces a current $j_a=Re(j^0_a+j^{2\omega}_a e^{2i\omega t})$. This non-linear current has two Fourier components at zero and twice the frequency of the applied external field, $j^0=\chi_{abc}\mathcal{E}_b\mathcal{E}^*_c$ and $j^{2\omega}=\chi_{abc}\mathcal{E}_b\mathcal{E}_c$. The response function $\chi_{abc}$, which can be expressed as $\chi_{abc}=-\epsilon_{adc}e^3 \tau D_{bd}/2(1+i\omega\tau)$, with $\epsilon_{adc}$ being the Levi-Civita tensor and $\tau$ the scattering time, explicitly contains the Berry curvature dipole \footnote{Asymmetric defect scattering and other extrinsic mechanisms can also contribute to a non-linear Hall current. These effects can be experimentally decoupled (see Ref.\citep{Ma2019}) from the Berry curvature dipole contribution we discuss here.} defined as:

\begin{equation}
D_{bd}=\int_{k}f_0(\partial_b \Omega_d)
\label{dipole}
\end{equation}
where $\partial_a=\partial_{k_a}$, $\int_k=\int d^dk/(2\pi)^d$ and $f_0$ is the Fermi-Dirac distribution. 
The Berry curvature $\Omega$ is defined, as usual, as the rotor of the Berry connection $A_a=-i\left\langle u_k | \partial_a | u_k\right\rangle$. Crystalline symmetries may constrain the dipole to be zero: in a two-dimensional crystal, the Berry curvature is a pseudoscalar, and therefore the dipole behaves as a pseudovector contained in the two-dimensional plane. As a result, for a Berry curvature dipole to be non-vanishing the system must have at most one mirror symmetry left. If one mirror is preserved the dipole will then be directed perpendicular to the mirror line.

  The wallpaper group of graphene is $p6mm$, generated by the point group $\mathcal{C}_{6v}$ and in-plane translations. The point group $\mathcal{C}_{6v}$ is comprised of a three-fold rotation $\mathcal{C}_3$, a two-fold rotation $\mathcal{C}_2$ and the mirror symmetry $\mathcal{M}_{x}$ \citep{bernevig}. The two-fold rotation $\mathcal{C}_2$, which coincides with inversion since spin-orbit coupling can be effectively neglected, is immediately broken assuming a staggered chemical potential between the two honeycomb sublattices (Semenoff mass \citep{semenoff}). The latter can be engineered by placing the graphene sheet on a substrate, for example lattice-matched h-BN \citep{giovannetti,hBNmanchester}.
Breaking inversion reduces the point group to $\mathcal{C}_{3v}$, gaps out the Dirac cones present at the high symmetry points $K$ and $K'$ in the Brillouin zone, and allows for a non-vanishing Berry curvature. To further lower the point group symmetry to $\mathcal{C}_{v}$ we apply a uniform uniaxial strain to the honeycomb lattice along one of the two main crystallographic directions. This uniaxial strain moves the massive Dirac cones away from the the high symmetry points along the $k_y=0$ line, as required by the combination of time-reversal $\Theta$ and mirror symmetry $\mathcal{M}_x$. The $k_y=0$ line is perpendicular to the mirror line and parallel to the dipole. 
Although the existence of a Berry curvature dipole is perfectly allowed from a symmetry perspective, the corresponding low-energy theory does not immediately entail the system with a non-vanishing Berry curvature dipole. 

To show this we first discuss the low-energy theory of strained graphene. The application of strain to the lattice deforms the primitive cell and the corresponding Brillouin zone (Fig. \ref{fig:schematic}b ) but also produces a difference between hopping amplitudes along the two main crystallographic directions. To first order in strain and momentum, the low-energy description of the system shows that the strain behaves effectively as a pseudo-gauge field \citep{Guinea2010}: $\mathcal{H}_1=v_F[(\xi k_x+\mathcal{A}_x)\sigma_x+ k_y\sigma_y]+\Delta\sigma_z/2$, where $v_F$ is the Fermi velocity, $\sigma_{x,y}$ are the Pauli matrices, $\xi$ is the valley degree of freedom, $\Delta$ is the Semenoff mass and $\mathcal{A}_x\propto \epsilon_{xx}-\epsilon_{yy}$ is the pseudo-gauge field generated by the strain. If a uniform uniaxial strain is applied to the system, the pure gauge term in $\mathcal{H}_1$ can be reabsorbed by performing a momentum shift: this corresponds to expanding the tight-binding Hamiltonian around the Dirac point for $\Delta=0$. However, as was shown in Ref. \citep{Juan2012} both through a quantum field theory and a tight-binding approach, the Fermi velocity becomes anisotropic when considering the momentum-strain coupling. The corresponding low-energy Hamiltonian which includes momentum-strain coupling terms is then of the form, $\mathcal{H}_2=\xi v_x k_x \sigma_x+ v_y k_y \sigma_y+\Delta\sigma_z/2$ where $v_x$ and $v_y$ are the two strain-dependent Fermi velocities. The Hamiltonian $\mathcal{H}_2$ possesses a finite Berry curvature but still produces a vanishing Berry curvature dipole unless an extra term $\propto\xi k_x\sigma_0$ is added. This can be seen by performing the integral in Eq. \ref{dipole} as in Ref. \citep{Sodemann2015}.

 We will now show that the warping of the Fermi surface generates a finite Berry curvature dipole regardless of the presence of a tilt mechanism. By taking into account the trigonal warping caused by the $k^2$ terms the effective Hamiltonian to second order in momentum and first order in strain is,
 
\begin{align}
\mathcal{H}_\mathrm{warped}=&\xi v_x k_x \sigma_x+v_y k_y \sigma_y+\dfrac{\Delta}{2}\sigma_z+\nonumber \\ 
&+(\lambda_1 k_y^2-\lambda_2 k_x^2)\sigma_x+2\xi\lambda_3 k_x k_y\sigma_y \label{Hwarp}
\end{align}

where $\lambda_{1,2,3}$ are the warping terms which in general are not equivalent when considering strain-momentum coupling terms $\mathcal{O}(\epsilon k^2)$. 
The Hamiltonian in Eq. \ref{Hwarp} can be derived from a low-energy expansion of the tight-binding model following similar steps to the ones in Ref. \citep{Juan2012} and the Supplemental Material \footnote{See Supplemental Material [url] for the derivation of the effective Hamiltonians of monolayer and bilayer graphene and additional Berry dipole plots for different gaps and strains.}.

 It is important to note that although Eq. \ref{Hwarp} has anisotropic warping terms, for a non zero-dipole to exist it is sufficient to consider anisotropic velocities and a $\mathcal{C}_3$ symmetric warping. We will make use of this result later on when discussing bilayer graphene.

\begin{figure}[tbp]
\includegraphics[width=1\columnwidth]{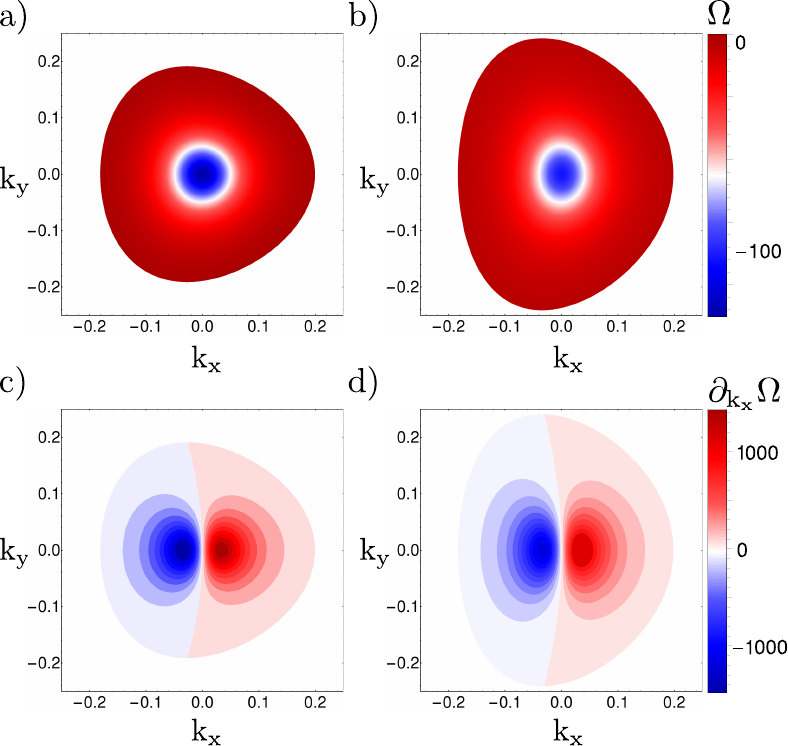}
\caption{Berry curvature $\Omega$ and dipole density $\partial_{k_x}\Omega$ of the conduction band corresponding to the Hamiltonian of Eq.~\ref{Hwarp} for unstrained (a,c) and strained (b,d) monolayer graphene. When strain is present the Fermi surface is deformed shifting the dipole moment from zero to a finite value. 
Momenta are measured in units of the inverse of the lattice constant $a$; the Berry curvature in units of $a^2$ while the dipole density in units of $a^3$.}
\label{fig:fig1}
\end{figure}

As shown in Fig.~\ref{fig:fig1}, when applying strain the warped surface deforms shifting the dipole moment from zero to a finite value leading to a Berry curvature dipole running along the zig-zag direction. Therefore when applying an electric field $E$ along the dipole direction a non-linear Hall current $\mathcal{O}(E^2)$ is generated along the armchair direction. We have computed the Berry curvature dipole \cite{Note2} generated by the two cones, which contribute equally to the overall response, by continuously changing the carrier density $n$ and different values of strain. At $5\%$ strain and for a Semenoff mass $\Delta=20 \,\mathrm{meV}$ a dipole $\mathrm{D_x}\approx 10^{-3}\,\mathrm{nm}$ is found at an electron density $n\approx 10^{10} \,\mathrm{cm^{-2}}$. 
This order of magnitude of $\mathrm{D_x}$ is comparable to the one predicted in Ref. \citep{Sodemann2015} for TMDs.

\paragraph{Berry Curvature Dipole in Strained Bilayer Graphene} 
Having established that uniaxially strained monolayer graphene possesses a sizable Berry curvature dipole,
we now show that the same effect persists also in bilayer graphene 
in the (AB) Bernal-stacked structure. Importantly, the Berry dipole in this material is boosted by over three order of magnitudes. Moreover, in bilayer graphene inversion symmetry breaking can be achieved with the application of an external electric field perpendicular to the layers. The electric field, in fact, generates a spectral gap $\Delta$ and lowers the point group symmetry from $D_{3d}$ to $C_{3v}$~\cite{McCann2013}. Notice that this inversion symmetry breaking mass can be experimentally tuned independent of the carrier density~\cite{Oostinga2008}. The additional application of a uniaxial strain reduces the point group to $C_v$ and yields a vanishing Berry dipole perpendicular to the mirror line. To prove the assertion above, we introduce a low-energy effective Hamiltonian \citep{Schrieffer1966}, valid for electron densities up to $n\approx 10^{13}\,\mathrm{cm^{-2}}$ \citep{McCann2006, Mucha-Kruczynski2010}, explicitly accounting for the effect of strain~\citep{Mucha-Kruczynski2011}, reading:

\begin{align}
{\mathcal H}_{b}= \left[ -\frac{1}{2m} (k_x^2 - k_y^2) + \xi v_3 k_x  + w\right]\sigma_x  \label{eq:Hdip} \\ \nonumber
-\left(\frac{1}{m}k_xk_y + \xi v_3 \right)\sigma_y + \frac{\Delta}{2}\sigma_z.
\end{align}
In the equation above, $\xi=\pm 1$ is the valley index, $v_3$ is the Fermi velocity related to the ``skew" hopping between the layers, whereas $m$ is an effective mass directly dependent on the interlayer coupling. Finally, $w=\mathcal{A}_{3}-\mathcal{A}_{0}$ is the strain term in the Hamiltonian which can be expressed in terms of the two pseudo-gauge fields $\mathcal{A}_{0,3}$.  

\begin{figure}[tbp]
\includegraphics[width=1.02\columnwidth]{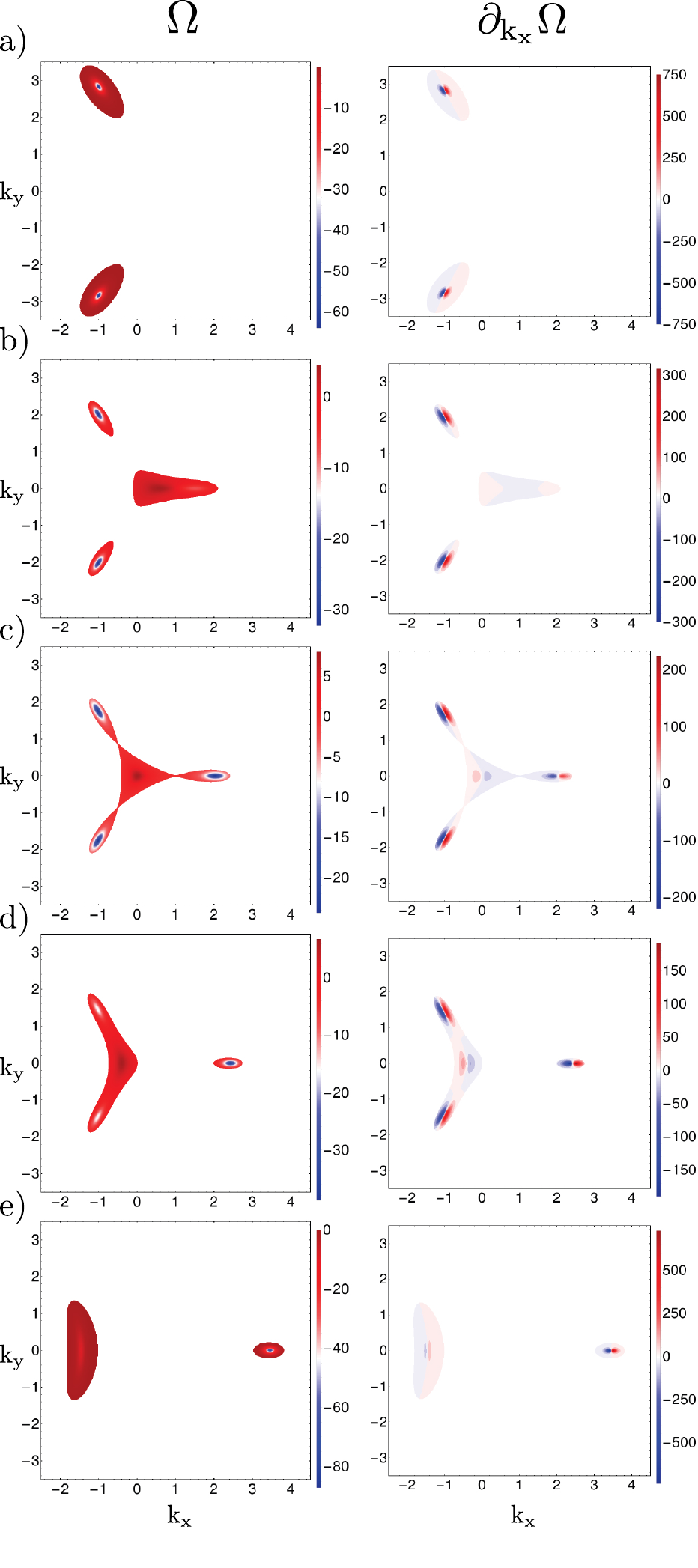}
\caption{Berry curvature $\Omega$ and dipole density $\partial_{k_x}\Omega$ of the conduction band corresponding to the low-energy Hamiltonian of Eq.~\ref{eq:Hdip} for different values of the strain: a) $w=-5~\varepsilon_L$, b) $w=-1~\varepsilon_L$, c) $w=0$, d) $w=1~\varepsilon_L$, e) $w=5~\varepsilon_L$. All plots are shown for the same carrier density fixed by placing the Fermi energy $E_F$ at the Lifschitz transition in the unstrained $w=0$ and gapped case (panel c). Momenta are measured in units of $\kappa_L$, the Berry curvature in units of $1/ \kappa_L^2$, and the  dipole density in units of $1/\kappa_L^3$.}
\label{fig:fig3}
\end{figure}

In the presence of inversion symmetry ($\Delta=0$), the strain-free ($w=0$) system features a Lifshitz transition~\cite{McCann2006} at energy $\varepsilon_L = m v_3^2/2$ $(\approx1 \,$ meV) where the Fermi surface splits from a single connected pocket into four different ones: the electronic dispersion consists of one central Dirac cone with $-\pi$ ($\pi$) Berry phase at the $K$ ($K^{\prime}$) point of the Brillouin zone and three ``leg" Dirac cones, each of which carries a $\pi$ ($-\pi$) Berry phase. Notice that the distance between the different cones defines a characteristic momentum $\kappa_L=m v_3/\hbar$ $(\approx 0.035 \,$ nm$^{-1}$).
The effect of the strain on the 
inversion symmetric system is twofold, as it moves the Dirac cones away their unstrained positions and it promotes changes in the topology of the Fermi surface by merging the cones together. For $- 1\leq w / \varepsilon_L \leq 3$ there are always four Dirac points, two on the $k_y=0$ line and the remaining two in a symmetric position with respect to it. At $w=-\varepsilon_L$ the Dirac cones on the invariant line merge and become gapped, giving rise to a local minimum in the dispersion relation which survives until $w\geq-9 \varepsilon_L$, after which it becomes a saddle point. For $w>3 \varepsilon_L$, instead, there are only the cones on the $k_y=0$ line and no other local minima. 
However, unless $\Delta\neq0$, the spectrum remains gapless and the Berry curvature is zero. As mentioned above, these gapless spectra become gapped with an externally applied electric field. Consequently, the Dirac points become local minima of the dispersion.

\begin{figure}[tbp]
\includegraphics[width=.9\columnwidth]{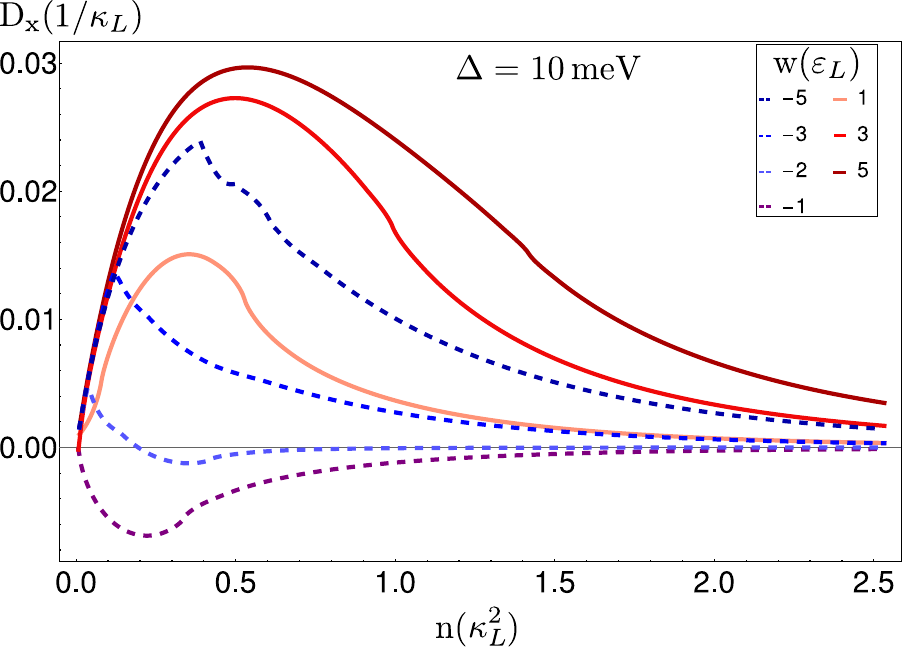}
\caption{Berry curvature dipole, measured in units of $1/\kappa_L$, in bilayer graphene for $\Delta=10\,\mathrm{meV}$ and various strains $w$ as a function of the electron density $n$ measured in units of $\kappa_L^2$ where $\kappa_L \simeq 0.035$~nm$^{-1}$. Densities of this order of magnitude have been experimentally reported in Ref.s \citep{Oostinga2008,Ohta,Novoselov}.}
\label{fig:fig4}
\end{figure}

The possible topologies of the Fermi surface induced by the strain are visible in Fig. \ref{fig:fig3}, 
where we plot the Berry curvature and the density of Berry curvature dipole at different values of $w$ and $\Delta\neq0$. We notice that the plots are symmetric for the exchange $k_y \rightarrow -k_y$, 
as dictated by the combination of time-reversal and mirror symmetry. 
The unstrained case is shown in Fig.~\ref{fig:fig3}(c). The threefold $\mathcal{C}_3$ rotation symmetry constrains the total Berry dipole to be zero in this case. 
It is crucial to notice, however, that while the central gapped Dirac cone has a vanishing Berry dipole, the three leg gapped Dirac cones have a non-zero dipole when taken by themselves. This is because each of the leg gapped Dirac cones can be described with an effective low-energy Hamiltonian of the form of Eq.~\ref{Hwarp} with $\lambda_1=\lambda_2=\lambda_3$. The perfect cancellation of these three non-zero contributions due to the threefold rotation symmetry  is lost in the presence of uniaxial strain. Moreover, in the presence of finite strain also the central gapped cone yields a non-zero contribution to the Berry dipole. Figs. \ref{fig:fig3}(a-b,d-e) actually suggest that when strain deforms the cones, a net Berry curvature dipole is generated. In order to verify this,  we have computed the Berry dipole as a function of the electron density for different values of $w$ [see Fig.\ref{fig:fig4}]. The mass has been chosen as $\Delta=10\varepsilon_L$. The behavior of the total Berry dipole has a richer structure as compared to the one for monolayer graphene: it indeed shows cusps and inflection points that are a different consequence of the different Lifshitz transitions and reflect the richer Fermiology of bilayer graphene. Furthermore, in this material it is possible to tune the strain in such a way that the sign of the dipole changes upon increasing the electron density with an external gating \cite{Oostinga2008}, thus inverting the direction of the transverse current or even suppressing it altogether. More importantly, at $w=-5\varepsilon_L$, which corresponds roughly to a $1\%$ strain~\citep{Mucha-Kruczynski2011}, and for a gap $\Delta=10 \,\mathrm{meV}$, we find for $n\approx 10^{11}\,\mathrm{cm^{-2}}$ \citep{Oostinga2008}  a  maximum dipole strength of $\mathrm{D_x}\approx 1\,\mathrm{nm}$. This value is three-order of magnitudes larger than that  of single-layer graphene, and is comparable to the Berry dipole experimentally found in bilayer WTe$_2$~\citep{Ma2019}. As shown in the Supplemental Material \citep{Note2}, even higher values on the tens of nanometer scale can be found by decreasing the inversion symmetry breaking mass.

\paragraph{Conclusions --} 
To wrap up, we have shown that non-vanishing Berry curvature dipoles can emerge even in the complete absence of spin-orbit coupling in two-dimensional Dirac materials as a result of the warping of the Fermi surface. We have in fact proved that, in the presence of substrate-induced and gate-induced band gaps respectively, uniaxially strained monolayer and Bernal-stacked bilayer graphene 
do possess sizeable Berry curvature dipoles.
In the bilayer structure, the Berry dipole is strongly enhanced and its value is comparable to the one experimentally observed in bilayer WTe$_2$. Since the warping of the Fermi surface is ubiquitous, we expect that our results apply to a large number of two-dimensional materials where strain engineering can be used to achieve the minimum symmetry constraints for a non-vanishing Berry curvature dipole. The corresponding non-linear Hall effect can thus be used as a way to directly probe the geometric properties of Bloch states in a large number of  time-reversal invariant two-dimensional materials.

\begin{acknowledgments}
We are indebted to our collaborators Sheng-Chin Ho, Ching-Hao Chang, and Tse-Ming Chen for related work on this subject \citep{chingao}. 
C.O. acknowledges support from a VIDI grant (Project 680-47-543) financed by the Netherlands Organization for Scientific Research (NWO).
\end{acknowledgments} 

\end{document}